\documentstyle[12pt]{article}

\input epsf.sty

\newcommand{\ri}{{\rm i}}
\newcommand{\re}{{\rm e}}

\newcommand{\cD}{{\cal D}}

\newcommand{\cI}{{\cal I}}
\newcommand{\cO}{{\cal O}}

\newcommand{\cN}{{\cal N}}

\newcommand{\OO}{\mathop{\otimes}}
\font\cmss=cmss12 
\def\1{\hbox{{1}\kern-.25em\hbox{l}}}
\def\bfZ{\relax{\hbox{\cmss Z\kern-.4em Z}}}

\begin{document}
\begin{titlepage}

\centerline{\large \bf Reconstruction of non-forward evolution kernels.}

\vspace{10mm}

\centerline{\bf A.V. Belitsky\footnote{Alexander von Humboldt Fellow.},
                D. M\"uller}

\vspace{5mm}

\centerline{\it Institut f\"ur Theoretische Physik, Universit\"at
                Regensburg}
\centerline{\it D-93040 Regensburg, Germany}

\vspace{10mm}

\centerline{\bf A. Freund}

\vspace{5mm}

\centerline{\it INFN, Sezione di Firenze, Largo E. Fermi 2}
\centerline{\it 50125 Firenze, Italy}

\vspace{25mm}

\centerline{\bf Abstract}

\hspace{0.8cm}

We develop a framework for the reconstruction of the non-forward kernels
which govern the evolution of twist-two distribution amplitudes and
off-forward parton distributions beyond leading order. It is based on the
knowledge of the special conformal symmetry breaking part induced by the
one-loop anomaly and conformal terms generated by forward next-to-leading
order splitting functions, and thus avoids an explicit two-loop calculation.
We demonstrate the formalism by applying it to the chiral odd and flavour
singlet parity odd sectors.

\vspace{5cm}


\vspace{0.5cm}


\end{titlepage}

\section{Introduction}

The possibility to access new characteristics of hadrons by means of the
deeply virtual Compton scattering \cite{MulRobGeyDitHor94,Ji96,Rad96} and
the hard diffractive hadron electroproduction \cite{Rad96,ColFraStr96}
processes has recently initiated a growing phenomenological interest in
the underlying non-perturbative elements --- the so-called off-forward
parton distributions (OFPD) --- which parametrize hadronic structure in
these reaction making use of the QCD factorization theorems. The main
feature of the processes is a non-zero skewedness, i.e.\ plus component,
$\Delta_+ = \eta$, of the $t$-channel momentum transfer $\Delta$.

One of the central issues which has been addressed in this context is the
description of the scaling violation phenomena in the cross section via the
evolution of the off-forward parton distributions. Since the OFPD is defined
as an expectation value of a non-local string operator, its $Q^2$-dependence
is governed by the renormalization of this operator Fourier transformed to
the momentum fraction space. Inasmuch as the generalized skewed kinematics
can be unambiguously restored \cite{GeyDitHorMulRob88} from the conventional
exclusive one, known as Efremov-Radyushkin-Brodsky-Lepage (ER-BL) region
$\eta = 1$, in what follows we deal formally with renormalization of the
ordinary distribution amplitudes which obey the ER-BL equation
\cite{EfrRad78,BroLep79}
\begin{equation}
\label{ER-BLequation}
\frac{d}{d \ln Q^2} \mbox{\boldmath$\phi$} (x, Q) =
\mbox{\boldmath$V$} \left(x, y | \alpha_s(Q) \right)
\OO^\re \mbox{\boldmath$\phi$} (y, Q) .
\end{equation}
Here we have introduced the exclusive convolution
\begin{eqnarray*}
\OO^\re \equiv \int_{0}^{1} dy ,
\end{eqnarray*}
to distinguish it from the inclusive one used later. Here
$\mbox{\boldmath$\phi$} = { {^Q\phi} \choose {^G\phi} }$ is the
two-dimensional vector and $\mbox{\boldmath$V$} (x, y | \alpha_s )$
is a $2 \times 2$-matrix of evolution kernels given by a series in
the coupling.

Several methods have been offered so far to solve the off-forward evolution
equation: numerical integration \cite{FraFreGuzStr98}, expansion of OFPD
w.r.t.\ an appropriate basis of polynomials \cite{Beletal97,ManPilWei97},
mapping to the forward case\footnote{This idea has earlier been applied
directly to the kernels in \cite{BelMul98a}.} \cite{Shu99,ShuGolBieMarRys99}
and solution in the configuration space \cite{BalBra89,KivMan99}.
The last three methods are based on the well-known fact that operators
with definite conformal spin do not mix in the one-loop approximation.
Beyond leading order the latter two methods can only be applied in the
formal conformal limit of QCD where the $\beta$-function is set equal
to zero and making use of the conformal subtraction scheme which removes
the special conformal symmetry breaking anomaly appearing in the minimal
subtraction scheme. Thus, we are only left with the former two methods
which allows for a successive improvement of the perturbative
approximations involved.

Up to now only orthogonal polynomial reconstruction method has allowed the
analysis of the scaling violation in the singlet sector in two-loop
approximation since only the anomalous dimensions required in the formalism
were available so far \cite{BelMul98a,Mue94,BelMul98b}. This was sufficient
to get a first insight into the NLO evolution corrections. However, in order
to have an access to the whole kinematical region, especially for small
$x$, $\eta$ and high precision handling of the $x \sim \eta$ domain, one
should look for a more efficient numerical treatment. This can be achieved
with the first method alluded to above. To do the direct numerical
integration of the evolution equation one needs the corresponding evolution
kernels whose Gegenbauer moments define the anomalous dimensions mentioned
earlier. For the time being the former were available at LO order only. The
flavour non-singlet ER-BL kernel ($\eta = 1$) was obtained in NLO by a
cumbersome analytical calculation \cite{Sar84,DitRad84,MikRad85}. As we
have mentioned above the continuation to $\eta \in [0,1]$ is a unique
procedure \cite{GeyDitHorMulRob88}, so that one can obtain in a simple way
the evolution kernels for OFPD. The goal of this paper is to outline a
method that allows one to construct the singlet ER-BL kernels by applying
conformal and supersymmetric constraints where the latter ones arise from
the $\cN = 1$ super Yang-Mills theory \cite{BukFroKurLip85,BelMulSch98}.
In this way we can avoid the direct diagrammatical calculation which would
be very difficult to handle otherwise since no appropriate technology
has been developed yet.

The derivation is based on the fairly well established structure of the
ER-BL kernel in NLO. Up to two-loop order we have
\begin{equation}
\mbox{\boldmath$V$} (x, y | \alpha_s )
= \frac{\alpha_s}{2\pi}
\mbox{\boldmath$V$}^{(0)} (x, y)
+ \left( \frac{\alpha_s}{2\pi} \right)^2
\mbox{\boldmath$V$}^{(1)} (x, y)
+ \cO (\alpha_s^3) ,
\end{equation}
with the purely diagonal LO kernel $\mbox{\boldmath$V$}^{(0)}$ in the
basis of Gegenbauer polynomials and NLO one separated in two parts:
$\mbox{\boldmath$V$}^{(1)} (x, y) = \mbox{\boldmath$V$}^{{\rm D}(1)}
(x, y) + \mbox{\boldmath$V$}^{{\rm ND}(1)} (x, y)$, with the diagonal
part which is entirely determined by the well-known forward DGLAP
splitting functions $\mbox{\boldmath$P$} (z)$ \cite{BelMul98a}
\begin{equation}
\label{PtoVDreduction}
{^{AB} V}^{\rm D} ( x, y )
= \int_{0}^{1} dz\,
\sum_{j = 0}^{\infty} \frac{w (y | \nu)}{N_j(\nu)}
C^{\nu (A)}_j (2x - 1) z^j\, {^{AB}\! P} (z)
C^{\nu (B)}_j (2y - 1),
\end{equation}
where $N_j(\nu)= 2^{ - 4 \nu + 1 }
\frac{ \Gamma^2 (\frac{1}{2}) \Gamma ( 2 \nu + j )}{\Gamma^2 (\nu)
( \nu + j ) j! }$ and $w (y | \nu) = (y \bar y)^{\nu-1/2}$ are the
normalization and weight factors, respectively. The non-diagonal piece
is fixed completely by the conformal constraints \cite{BelMul98b}
\begin{equation}
\label{NDkernel}
\mbox{\boldmath$V$}^{{\rm ND}(1)} (x, y)
= - ( \cI - \cD )\,
\left\{
\mbox{\boldmath$\dot V$} \OO^\re
\left(
\mbox{\boldmath$V$}^{(0)} + \frac{\beta_0}{2}\, \1
\right)
+
\left[
\mbox{\boldmath$g$} \OO^\re_{,} \mbox{\boldmath$V$}^{(0)}
\right]_-
\right\} (x, y) ,
\end{equation}
in terms of
\begin{equation}
\mbox{\boldmath$V$}^{(0)}
=
\left(
\begin{array}{rr}
C_F\, {^{QQ}V^{(0)}}  &  2 T_F N_f\, {^{QG}V^{(0)}} \\
C_F\, {^{GQ}V^{(0)}}  &  C_A\, {^{GG}V^{(0)}}
\end{array}
\right) ,
\quad
\mbox{\boldmath$g$}
=
\left(
\begin{array}{rr}
C_F\, {^{QQ}g}  &  0            \\
C_F\, {^{GQ}g}  &  C_A\, {^{GG}g}
\end{array}
\right) ,
\end{equation}
the ER-BL kernels at LO and the special conformal symmetry breaking
matrix $\mbox{\boldmath$g$}$. Here $\beta_0 = \frac{4}{3} T_F N_f
- \frac{11}{3} C_A$ is the first expansion coefficient of the QCD
$\beta$-function. In the parity odd sector the dotted kernel,
$\mbox{\boldmath$\dot V$}$, is simply given by a logarithmic
modification of the $\mbox{\boldmath$V$}^{(0)}$. Due to subtleties,
appearing in the parity even case \cite{BelMul98a}, we deal here, for
the sake of simplicity, only with the parity odd and transversity
sectors.

The main problem is thus to restore the diagonal part of the
NLO kernels. Since the use of Eq.\ (\ref{PtoVDreduction}) beyond
LO is extremely complicated in practice, we are forced to look for
other solutions. It turns out that the bulk of contributions in the
ER-BL kernel can be deduced by going to the forward limit making use
of the reduction
\begin{eqnarray}
\label{SingletLimit}
\mbox{\boldmath$P$} (z)
= {\rm LIM}\, \mbox{\boldmath$V$} (x, y)
\equiv \lim_{\tau\to 0} \frac{1}{|\tau|}
\left(
\begin{array}{rr}
{^{QQ} V}
&
\frac{1}{\tau}{^{QG} V}
\\
\frac{\tau}{z} {^{GQ} V}
&
\frac{1}{z}{^{GG} V}
\end{array}
\right)^{\rm ext}
\left( \frac{z}{\tau}, \frac{1}{\tau} \right) .
\end{eqnarray}
Then the difference\footnote{Here $\mbox{\boldmath$V$}^{\rm ND}$
is understood without the $(\cI - \cD)$-projector.}
\begin{eqnarray*}
\mbox{\boldmath$P$} (z) - \mbox{\boldmath$P$}^{\rm cross-ladder} (z)
- {\rm LIM}\, \mbox{\boldmath$V$}^{\rm ND} (x, y)
\end{eqnarray*}
can be represented in terms of inclusive convolutions of simple splitting
functions and the back transformation to the exclusive kinematics is
trivial. The contributions of the purely diagonal cross-ladder diagrams
$\mbox{\boldmath$V$}^{\rm cross-ladder} (z)$ can be found from the known
$QQ$ sector \cite{Sar84,DitRad84,MikRad85} exploiting the $\cN = 1$
supersymmetric constraints.

The paper is organized as follows. In the next section we analyze the
structure of the known flavour non-singlet ER-BL kernel and state
the benchmarks of the formalism. The structure observed will give
us a guideline for construction of all other kernels: quark chiral
odd sector is considered in Section 3 and parity odd flavour singlet
one is discussed in Section 4. Finally, we give our conclusions and an
outlook .

\section{Structure of ER-BL kernel in non-singlet sector.}

It is very instructive to demonstrate the machinery in the simplest
case of non-singlet sector. Since the explicit two-loop calculation
is available \cite{Sar84,DitRad84,MikRad85} the direct comparison can
be made. The NLO $QQ$-kernel can be decomposed in colour structures
as\footnote{We omit the superscript $QQ$ later in this section.}
\begin{eqnarray}
\label{kernel-NS}
V (x, y | \alpha_s)
&=& \frac{\alpha_s}{2\pi}\, C_F V^{(0)}(x, y) \nonumber\\
&+& \left( \frac{\alpha_s}{2\pi} \right)^2
C_F
\left[
C_F V_F (x, y)
- \frac{\beta_0}{2} V_\beta (x, y)
- \left( C_F - \frac{C_A}{2} \right) V_G (x, y)
\right]_+ \nonumber\\
&+& \cO \left( \alpha_s^3 \right) ,
\end{eqnarray}
with the LO kernel $V^{(0)} (x, y) = \left[ v(x,y)\right]_+$, where
\begin{equation}
v(x,y) = \theta(y - x) f (x, y)
+ \theta(x - y) \overline f (x, y),
\quad\mbox{and}\quad
f (x, y) = \frac{x}{y} \left( 1 + \frac{1}{y - x} \right) .
\end{equation}
The shorthand notations $\bar x = 1 - x$ and $\overline f = f
(\bar x, \bar y)$ are used throughout the paper. The ``+''-prescription
is conventionally defined by
\begin{eqnarray*}
\left[ V (x, y) \right]_+ = V (x, y)
- \delta(x - y) \int_0^1 dz\, V (z, y).
\end{eqnarray*}

Let us now recall a few properties of the kernel that are useful for
the following considerations. Due to absence of the conformal symmetry
breaking counterterms at leading order for the renormalization of the
composite operators with total derivatives, one can use its consequences
to fix the eigenfunctions which turn out to be the Gegenbauer
polynomials $C_j^{3/2} (2x - 1)$ \cite{EfrRad78,BroLep79}. Thus, the
LO kernel is symmetric with respect to the weight function $x \bar x$:
$y \bar y V^{(0)} (x, y) = x \bar x V^{(0)} (y, x)$. Its eigenvalues are
given by the anomalous dimensions appeared in the analysis of deep
inelastic scattering. Thus, it is not surprising that a simple limit
already mentioned in Eq.\ (\ref{SingletLimit}) gives us the DGLAP kernel
\cite{GeyDitHorMulRob88}:
\begin{eqnarray}
\label{LIM-NS}
P(z) = {\rm LIM}\, V (x, y)
\equiv \lim_{\tau \to 0}
\frac{1}{|\tau|}
V^{\rm ext}\left(\frac{z}{\tau},\frac{1}{\tau}\right).
\end{eqnarray}
To perform this limit, we have to extend at first the ER-BL kernel,
originally defined in the domain $0 \leq x, y \leq 1$, to the whole
region $x, y \in ( - \infty, \infty )$ by a unique procedure which is
given in practice by the replacement, e.g.\ at leading order, of the
$\theta$-function by
\begin{eqnarray}
\label{Extension}
\theta (y - x)
\to
\theta \left( 1 - \frac{x}{y} \right)
\theta \left( \frac{x}{y} \right)
\mbox{sign} (y).
\end{eqnarray}
If a kernel is diagonal in the ER-BL representation, we can restore it
from the known DGLAP kernel by the integral transformation
(\ref{PtoVDreduction}). Because of branch cuts appearing in the
convolutions of the NLO terms with the transformation kernel, it is highly
nontrivial to handle the inverse reduction to the exclusive kinematics.

At NLO the kernel (\ref{kernel-NS}) contains besides a pure diagonal part
with respect to the Gegenbauer polynomials also a non-diagonal part
located in $V_F (x, y)$ and $V_\beta (x, y)$. These parts are predicted by
conformal constraints (see $QQ$-entry of Eq.\ (\ref{NDkernel})) and are
fixed by the one-loop special conformal anomaly kernels
\cite{BelMul98a,Mue94,BelMul98b}:
\begin{equation}
\label{def-dV-NS}
\dot v (x, y)
= \theta(y - x) f (x, y) \ln \frac{x}{y}
+ \left\{ x \to \bar x \atop y \to \bar y \right\},
\quad
g (x, y)
= - \theta(y - x)
\frac{ \ln \left( 1 - \frac{x}{y} \right) }{y - x}
+ \left\{ x \to \bar x \atop y \to \bar y \right\} .
\end{equation}

Let us now analyze in detail the contributions to the NLO kernel from
different colour structures. The expressions for $C_F^2$ terms
arise from Feynman diagrams containing quark self-energy insertions
and ladder graphs\footnote{For simplicity we imply the diagrams
in the light-cone gauge \cite{Sar84,DitRad84}.}. In order to subtract
the ultraviolet (UV) divergences in subgraphs it requires the LO
renormalization of the composite operator to which these lines are
attached to. The explicit calculation gives \cite{Sar84,DitRad84,MikRad85}
\begin{eqnarray}
\label{kernel-NS-CFa}
V_F (x, y)
&=& \theta (y - x)
\Bigg\{ \left( \frac{4}{3} - 2 \zeta (2) \right) f
+ 3 \frac{x}{y}
- \left( \frac{3}{2} f - \frac{x}{2 \bar y} \right) \ln \frac{x}{y}
- ( f - \overline f ) \ln \frac{x}{y}
\ln \left( 1 - \frac{x}{y} \right)
\nonumber \\
&+&
\left( f + \frac{x}{2 \bar y} \right) \ln^2 \frac{x}{y} \Bigg\}
- \frac{x}{2\bar y} \ln x \left( 1 + \ln x - 2 \ln \bar x \right)
+ \left\{ {x \to \bar x \atop y \to \bar y } \right\}.
\end{eqnarray}
Making use of the known non-diagonal part (\ref{NDkernel}), $V_F$ can be
represented up to a pure diagonal term, denoted as $D_F(x,y)$, by the
convolution
\begin{equation}
V_F (x, y) =
- \left( \dot{v} \OO^\re v
+ g \OO^\re v - v \OO^\re g \right) (x, y)
+ D_F (x, y).
\end{equation}
To find an appropriate representation of this missing diagonal
element we first take the forward limit. Since the forward limit
of the convolution is\footnote{We remind as well that $[A]_+ \OO [B]_+
= [C]_+$.}
\begin{equation}
{\rm LIM}\, \left\{ [\dot v]_+ \OO^\re [v]_+ \right\} =
\left\{ {\rm LIM}\, [\dot v]_+ \right\}
\OO^\ri \left\{ {\rm LIM}\, [v]_+ \right\} ,
\end{equation}
where we have introduced the inclusive convolution
\begin{eqnarray*}
P_1 (z) \OO^\ri P_2 (z)
\equiv \int_0^1 dx \int_0^1 dy \delta( z - xy ) P_1 (x) P_2 (y) ,
\end{eqnarray*}
and the commutator $g \OO V^{(0)} - V^{(0)} \OO g$ drops out in the
forward limit, we obtain
\begin{eqnarray}
{\rm LIM}\, V_F (x, y)
= - \dot p \OO^\ri p
+ {\rm LIM}\, D_F (x,y) ,
\end{eqnarray}
where $\dot p = {\rm LIM}\, \dot v = p(z) \ln z + 1 - z$ and
$p (z) = {\rm LIM}\, v (x, y) = (1 + z^2)/(1 - z)$. The comparison
of ${\rm LIM}\, V_F (x, y)$ with the corresponding part of the DGLAP
kernel \cite{CurFurPet80}
\begin{eqnarray}
P_F (z)
&=& \left\{ \frac{4}{3} - 2 \zeta (2)
- \frac{3}{2} \ln z + \ln^2 z - 2\ln z \ln(1 - z)
\right\} p(z) \nonumber\\
&+& 1 - z + \frac{1 - 3 z}{2} \ln z  - \frac{1 + z}{2} \ln^2 z ,
\end{eqnarray}
yields the result in which all double log terms are contained
in the convolution $\dot p \OO p$ and, therefore, only
single logs survive in $D_F (z) = {\rm LIM} D_F (x, y)$:
\begin{eqnarray}
D_F(z) &=& P_F + \dot p \OO^\ri p \nonumber\\
&=& - \frac{1}{2} p^a (-z) \ln z
- p^a (z) \left\{ \ln z - 2 \ln (1 - z) - \frac{1}{2} \right\}
- \frac{5}{12} p (z) .
\end{eqnarray}
Here we have introduced for convenience the kernel $p^a (z) = 1 - z$.
The next important point is that the remaining log terms can be
represented as convolutions of $p^a$ and $p$. Thus, we have finally
\begin{eqnarray}
D_F (z)
= \frac{1}{2} p^a \OO^\ri \left\{ 2\, p + p^a \right\} (z)
+ \frac{1}{12} p(z) + \frac{5}{2} p^a(z) .
\end{eqnarray}
Since $D_F (x, y)$ is by definition diagonal, the extension of $D_F (z)$
towards the ER-BL kinematics is trivial:
\begin{eqnarray}
\label{DF-QQ}
D_F (z) \to D_F (x, y)
= \frac{1}{2} v^a \OO^\re
\left( 2\, v + v^a \right) (x, y)
+ \frac{1}{12} v (x, y) + \frac{5}{2} v^a (x, y),
\end{eqnarray}
where a new diagonal element is $v^a (x,y) = \theta(y - x)
\frac{x}{y} + \theta(x - y) \frac{\bar x}{\bar y}$. Evaluating
the convolutions one can establish the equivalence of our prediction
with Eq.\ (\ref{kernel-NS-CFa}).

Next, the Feynman diagrams containing vertex and self-energy corrections
provide  $V_\beta$ proportional to $\beta_0$. Its off-diagonal part is
induced by the renormalization of the coupling and is contained in the
dotted kernel (\ref{def-dV-NS})
\begin{equation}
V_\beta (x, y) = \dot v (x, y) + D_\beta (x, y) .
\end{equation}
The remaining diagonal piece, $D_\beta$, is deduced from the known NLO
DGLAP kernel
\cite{CurFurPet80}
\begin{equation}
\label{kernelP-NS-beta}
P_\beta (z) = \frac{5}{3} p (z) + p^a (z) + \dot p (z)
\end{equation}
by going to the forward kinematics and restoring then the missed
contributions from it. Thus,
\begin{equation}
\label{Dbeta-QQ}
D_\beta (x, y) = \frac{5}{3} v (x, y) + v^a (x, y) .
\end{equation}
Indeed, the final result coincides with \cite{Sar84,DitRad84,MikRad85}.

Finally, we come to the contribution which mainly originates from the
crossed ladder diagram proportional to $(C_F - C_A/2)$:
\begin{eqnarray}
\label{kernel-NS-CAa}
V_G (x, y)
= 2 v^a (x, y) + \frac{4}{3} v (x, y)
+ \left( G (x, y)
+ \left\{ x \to \bar x \atop y \to \bar y \right\} \right).
\end{eqnarray}
Since this diagram has no UV divergent subgraph and thus requires no
subtraction, its contribution has to be diagonal w.r.t.\ the Gegenbauer
polynomials. This is obvious for the first two terms appearing in
Eq.\ (\ref{kernel-NS-CAa}). The function\footnote{We have slightly
changed the original definition given in \cite{Sar84,DitRad84} by
$G (x, y) + 2 \theta (y - x) \overline f \ln y \ln \bar x \to G (x, y)$.}
$G(x,y)$ contains in the unusual $\theta (y - \bar x)$-structure
the mixing between quarks and antiquarks
\begin{equation}
\label{kernel-NS-G}
G (x, y) = \theta (y - x) H (x, y)
+ \theta (y - \bar x) \overline H (x, y),
\end{equation}
with
\begin{eqnarray}
H (x, y)
&=& 2 \left[ \overline f
\left( {\rm Li}_2 (\bar x) + \ln y \ln \bar x \right)
- f\, {\rm Li}_2 (\bar y) \right] , \\
\overline H (x, y)
&=& 2 \left[ ( f - \overline f )
\left( {\rm Li}_2 \left( 1 - \frac{x}{y} \right)
+ \frac{1}{2} \ln^2 y \right)
+ f \left( {\rm Li}_2 (\bar y) - {\rm Li}_2 (x)- \ln y \ln x \right)
\right] ,
\end{eqnarray}
where ${\rm Li}_2$ is the dilogarithm. It can be easily checked that
the $G$-contribution (not the terms $H$ and $\overline H$ separately)
is symmetrical w.r.t.\ the weight $x \bar x$. Performing the limit
(\ref{LIM-NS}) we obtain the following correspondence with the
non-singlet DGLAP kernel \cite{MulRobGeyDitHor94}:
\begin{equation}
\label{G-NS}
G(z) \equiv {\rm LIM} G (x, y)
= \theta(z) \theta(1 - z) H (z)
+ \theta(- z) \theta(1 + z) \overline H (z) ,
\end{equation}
where
\begin{eqnarray}
\label{LIM-H-NS}
H (z) &\equiv& {\rm LIM}\, H (x, y)
= p(z) \left( \ln^2 z - 2 \zeta (2) \right) + T (z) , \\
\label{LIM-bH-NS}
\overline H (z) &\equiv& {\rm LIM}\, \overline H (x, y)
= 2 p(z) S_2(- z) + T (-z) .
\end{eqnarray}
Here $S_2 (z) = \int_{z/(1+z)}^{1/(1+z)} \frac{dx}{x}\ln\frac{1 - x}{x}$
and $T (z) = 2 (1 + z) \ln z + 4(1 - z)$. We should emphasize that
this structure of $G$ is the most general, especially, it is the
only contribution that contains Spence functions. In the forward
limit we obtain therefore a typical combinations given in
(\ref{LIM-H-NS}) and (\ref{LIM-bH-NS}), which can be found in all
other channels as well. This observation provides us with a hint for the
construction of all singlet $G$ kernels in the ER-BL representation.
For completeness, we give the corresponding part of the DGLAP kernel
\cite{CurFurPet80}
\begin{eqnarray}
\label{kernelP-NS-CA}
P_G (z) = {\rm LIM}\, V_G (x, y)
= 2 p^a (z) + \frac{4}{3} p(z)+ G(z),
\end{eqnarray}
and $G(z)$ defined above in Eqs.\ (\ref{G-NS})-(\ref{LIM-bH-NS}).

Recapitulating the results obtained in this section, we have observed a
rather simple structure of the non-singlet NLO kernel in the $QQ$-channel.
Up to the diagonal $G$-function, which is in fact the only new element in
the two-loop approximation, we can represent all other terms by a simple
convolution of LO kernels already known. It is not accidental but
a mere consequence of the topology of contributing Feynman graphs
at $\cO (\alpha_s^2)$. Thus, we anticipate the same feature to appear
in all other channels as well.

\section{Quark kernel in chiral odd sector.}

After we have outlined and tested in the preceding section our
formalism, we can apply it to the previously unknown transversity
two-loop ER-BL kernel. We decompose the transversity kernel analogous
to the chiral even case (\ref{kernel-NS}). We also use the same
decomposition for the DGLAP kernels \cite{Vog97}. The leading order
kernel is
\begin{equation}
\label{kernel-tr-0}
V^{(0)T} (x, y)
= \left[ v^b (x, y) \right]_+ - \frac{1}{2} \delta(x - y),
\end{equation}
with
\begin{equation}
v^b (x, y) = \theta(y - x) f^b (x, y)
+ \theta(x - y) \overline f^b (x, y),
\quad\mbox{and}\quad
f^b (x, y) = \frac{x}{y} \frac{1}{y - x}.
\end{equation}

The non-diagonal part has been analyzed in Ref.\ \cite{BelMul98b} and
is completely analogous to the chiral even case discussed above. Thus,
\begin{eqnarray}
V_F^T (x, y)
= - \left\{
\left[ \dot v^b \right]_+ \OO^\re V^{(0)T}
+ \left[ g \right]_+ \OO^\re V^{(0)T}
- V^{(0)T} \OO^\re \left[ g \right]_+
\right\} (x, y) + D_F^T (x, y),
\end{eqnarray}
where $\dot v^b$ is obtained from Eq.\ (\ref{def-dV-NS}) by
replacing $f$ by $f^b$ and the $g$ kernel is the same as in Eq.\
(\ref{def-dV-NS}). Taking the forward limit of $V_F^T (x, y)$
and comparing it with the known result for the DGLAP kernel in
Ref.\ \cite{Vog97}, we find the following trivial representation of
the remaining diagonal part
\begin{eqnarray}
D_F^T (x, y)=
- \frac{2}{3} \left[ v^b (x, y) \right]_+
- \frac{19}{24} \delta(x - y).
\end{eqnarray}

There is essentially no extra work required to find the contribution
proportional to the $\beta_0$-function, since it can be easily traced
from the DGLAP kernel \cite{Vog97} to be
\begin{eqnarray}
\label{kernel-tr-beta}
V_\beta^T (x, y)
= \frac{5}{3} \left[ v^b (x, y) \right]_+
+ \left[ \dot v^b (x, y) \right]_+
- \frac{13}{12} \delta(x - y).
\end{eqnarray}

The case of the $G$ function is easy to handle as well. If we replace
$f$ by $f^b$ in the definition (\ref{kernel-NS-G}), we obtain the
diagonal $G^T (x, y)$ kernel. Taking the forward limit and comparing
it with the DGLAP kernel, we immediately find the remaining
$\delta$-function contribution, so that the whole result reads
\begin{eqnarray}
\label{kernel-tr-CAa}
V^T_G (x, y)
= \left[
G^T (x, y)
+ \left\{ {x \to \bar x \atop y \to \bar y } \right\}
\right]_+
- \frac{19}{6} \delta(x - y).
\end{eqnarray}
This completes the discussion of the quark chiral-odd channel.

\section{Flavour singlet parity odd sector.}

Let us now address the flavour singlet parity odd sector responsible
for the evolution of axial-vector distribution amplitudes. For even parity
there are few subtleties, which will be discussed elsewhere. Here
we would only like to note that in the latter case a direct leading order
calculation provides a result that suffers for the mixed channel from
off-diagonal matrix elements in the unphysical sector. Although the
improved result has been found in Ref.\ \cite{BelMul98a}, it still
remains a difficult task to find an appropriate representation for the
dotted kernels and the two-loop $G$ functions.

Making use of the known non-diagonal part of the ER-BL kernel
(\ref{NDkernel}), the whole NLO result in the axial-vector case reads
\begin{equation}
\label{pred-Sing}
\mbox{\boldmath$V$}^{(1) A}
= - \mbox{\boldmath$\dot V$}^{A} \OO^\re
\left(
\mbox{\boldmath$V$}^{(0)A} + \frac{\beta_0}{2}\, \1
\right)
- \mbox{\boldmath$g$} \OO^\re \mbox{\boldmath$V$}^{(0)A}
+ \mbox{\boldmath$V$}^{(0)A} \OO^\re \mbox{\boldmath$g$}
+ \mbox{\boldmath$D$}^{A} + \mbox{\boldmath$G$}^{A},
\end{equation}
where the kernels $\mbox{\boldmath$D$}^{A} (x, y)$ and
$\mbox{\boldmath$G$}^{A} (x, y)$ are purely diagonal. Here the matrix
of the LO kernels is given in a compact form by
\begin{eqnarray}
\mbox{\boldmath$V$}^{(0)A} (x, y)
=
\left(
\begin{array}{ll}
C_F \left[ {^{QQ} v} (x, y) \right]_+
& - 2 T_F N_f \, {^{QG} v^a} (x, y) \\
C_F\, {^{GQ} v^a} (x, y)
& C_A \left[ {^{GG} v^A} (x, y) \right]_+
- \frac{\beta_0}{2} \delta(x - y)
\end{array}
\right) ,
\end{eqnarray}
where
${^{QQ} v} \equiv {^{QQ} v^a} + {^{QQ} v^b}$ and
${^{GG} v^A} \equiv 2\, {^{GG} v^a} + {^{GG} v^b}$.
The general structure of the functions $v^i$ is
\begin{equation}
{^{AB} v^i}(x, y)
= \theta(y - x) {^{AB}\! f^i}(x, y)
\pm \left\{ {x \to \bar x \atop y \to \bar y } \right\}
\quad
\mbox{for}
\quad
\left\{ {A = B \atop A \not = B } \right. ,
\end{equation}
with
\begin{equation}
\left\{{ {^{AB}\! f^a} \atop {^{AB}\! f^b} }\right\}
= \frac{ x^{\nu(A) - 1/2}}{y^{\nu(B) - 1/2}}
\left\{ { 1 \atop \frac{1}{y - x} } \right\} .
\end{equation}
The index $\nu(A)$ coincides with the index of Gegenbauer polynomials
in the corresponding channel, i.e.\ $\nu(Q) = 3/2$ and $\nu(G) = 5/2$.
The dotted kernels involved in the definition (\ref{pred-Sing}) can
simply be obtained by differentiating LO results w.r.t.\ the index $\nu$
which gives rise to the additional $\ln(x/y)$-multiplier in front of the
former
\begin{equation}
\mbox{\boldmath$\dot V$}^{(0)A} (x, y)
=
\left(
\begin{array}{ll}
C_F \left[ {^{QQ} \dot v} (x, y) \right]_+
& - 2 T_F N_f {^{QG} \dot v}^a (x, y) \\
C_F {^{GQ} \dot v}^a (x, y)
&
C_A \left[ {^{GG} \dot v}^A (x, y) \right]_+
\end{array}
\right) ,
\end{equation}
with the matrix elements
\begin{equation}
{^{AB} \dot v} (x, y) =
\theta(y - x) {^{AB}\! f} (x, y) \ln \frac{x}{y}
\pm \left\{ {x \to \bar x \atop y \to \bar y } \right\} ,
\quad
\mbox{for}
\quad
\left\{ { A = B \atop A \not= B } \right. .
\end{equation}
Note that for $A = B$ the dotted kernels are defined with the
``+''-prescription. The $\mbox{\boldmath$g$}$ function is given by
\begin{eqnarray}
\label{set-g-kernels}
\mbox{\boldmath$g$} (x, y) =
\theta(y - x)
\left(
\begin{array}{cc}
- C_F \left[ \frac{ \ln \left( 1 - \frac{x}{y} \right) }{y - x} \right]_+
& 0 \\
C_F \frac{x}{y}
& - C_A\left[ \frac{ \ln \left( 1 - \frac{x}{y} \right) }{y - x} \right]_+
\end{array}
\right)
\pm
\left\{ x \to \bar x \atop y \to \bar y \right\},
\end{eqnarray}
with ($-$) $+$ sign corresponding to (non-) diagonal elements.
Note, that we have used the property $(\cI - \cD) \ln (1 - \frac{x}{y})
= - (\cI - \cD) \frac{x}{y}$ for the element of $GQ$-channel to make
contact with the results of Ref.\ \cite{BelMul98b}.

Next we construct the diagonal $\mbox{\boldmath$G$} (x, y)$ kernel. At
first glance one would naively expect that one can obtain these kernels
by only inserting appropriate ${^{AB}\! f}$ functions in the
definition (\ref{kernel-NS-G}), so that the symmetry properties of the
$f$ functions w.r.t.\ the weight induce then the desired symmetry of
the ${^{AB} G}$ functions. Unfortunately, the symmetry is not sufficient
for the diagonal form of the $\mbox{\boldmath$G$} (x, y)$ kernel. To
ensure the diagonality, we have to add terms containing single logs and
rational functions. Let us define the matrix
\begin{equation}
\label{G-kernel-odd}
\mbox{\boldmath$G$}^A (x, y)
= - \frac{1}{2}
\left(
\begin{array}{cc}
2 C_F \left( C_F - \frac{C_A}{2} \right)
\left[ {^{QQ} G}^A (x, y) \right]_+
&
2 C_A T_F N_f \, {^{QG} G}^A (x, y)
\\
C_F C_A \, {^{GQ} G}^A (x, y)
&
C_A^2 \left[ {^{GG} G}^A (x, y) \right]_+
\end{array}
\right) ,
\end{equation}
with the following general structure
\begin{equation}
{^{AB} G}^A (x, y)
= \theta (y - x)
\left( {^{AB}\! H}^A + \Delta{^{AB}\! H}^A \right) (x, y)
+ \theta (y - \bar x)
\left( {^{AB} \overline H}^A
+ \Delta{^{AB} \overline H}^A \right) (x, y) .
\end{equation}
Here analogous to the non-singlet case we set
\begin{eqnarray}
\label{kernel-S-H}
{^{AB} H}^A (x, y)
\!\!&=&\!\! 2 \left[ \pm {^{AB} \overline f}^A
\left( {\rm Li}_2( \bar x ) + \ln y \ln \bar x \right)
- {^{AB}\! f}^A\, {\rm Li}_2( \bar y ) \right], \\
\label{kernel-S-bH}
{^{AB} \overline{H}}^A (x, y)
\!\!&=&\!\! 2 \left[
\left( {^{AB}\! f}^A \mp {^{AB} \overline f}^A \right)
\left( {\rm Li}_2 \left( 1 - \frac{x}{y} \right)
+ \frac{1}{2} \ln^2 y \right)
+ {^{AB}\! f}^A \left( {\rm Li}_2 ( \bar y )
- {\rm Li}_2 (x) - \ln y \ln x \right) \right],
\nonumber\\
\end{eqnarray}
where the upper (lower) sign corresponds to the $A = B$ ($A \not= B$)
channels. An explicit use of the reduction $P \to V^{\rm D}$ procedure
(\ref{PtoVDreduction}) to restore the $\Delta H$ contributions is rather
involved due to complexity of the integrand function. Rather we have
succeeded to deduce them using different arguments. Since the crossed
ladder diagrams have no UV divergent subgraphs the kernels ${^{AB} G}$
in different channels are related in a scheme independent way by
supersymmetry and conformal covariance of $\cN = 1$ super Yang-Mills
theory \cite{BelMulSch98}. Employing these symmetries we
restore\footnote{The details will be presented elsewhere.} all
necessary terms in a straightforward manner to be
\begin{eqnarray}
\Delta{^{QQ} H}^A (x, y)
&=& \Delta{^{QQ} \overline H}^A (x, y) = 0,
\\
\Delta {^{QG} H}^A (x, y)
&=& 2 \frac{\bar x}{y \bar y} \ln\bar x - 2 \frac{x}{y \bar y} \ln y,
\quad
\Delta{^{QG} \overline H}^A (x, y)
= 2 \frac{x}{y \bar y} \ln x - 2 \frac{\bar x}{y \bar y} \ln y,
\\
\Delta{^{GQ} H}^A (x, y)
&=& 2 \frac{x \bar x}{y} \ln\bar x - 2 \frac{x \bar x}{\bar y} \ln y,
\quad
\Delta{^{GQ} \overline H}^A (x, y)
= - 2 \frac{x \bar x}{y} \ln x + 2 \frac{x \bar x}{\bar y} \ln y,
\\
\Delta{^{GG} H}^A (x, y)
&=& \frac{x^2}{y^2} - \frac{1 + (x - y)^2}{y^2 \bar y^2}
- 2 \frac{x \bar x}{\bar y^2} \ln \frac{x}{y}
+ 2\frac{\bar x (\bar x - x)}{y \bar y} \ln\bar x
- 2 \frac{x(\bar x - x)}{y \bar y} \ln y,
\\
\Delta{^{GG} \overline H}^A (x, y)
&=& 2 \frac{x}{y^2} - \frac{x^2}{\bar y ^2}
+ 2 \frac{1 - x \bar x}{y \bar y^2}
+ 2 \frac{x \bar x}{y^2} \ln\frac{\bar x}{x}
+ 2 \frac{(x + \bar y) \bar x}{y \bar y^2} \ln \frac{x}{y}
- 2 \frac{1 - x \bar x}{y \bar y} \ln x
+ 6 \frac{x \bar x}{y \bar y} \ln y . \nonumber
\end{eqnarray}

Finally, we have to extract the remaining diagonal piece
$\mbox{\boldmath$D$}^A$ of $\mbox{\boldmath$V$}^A$ in the forward
limit (\ref{SingletLimit}) from the known DGLAP kernel
$\mbox{\boldmath$P$}^A$ \cite{MerNeeVog96}. We take into account
the underlying symmetry of the singlet parton distributions to map
the antiparticle contribution, i.e. $z < 0$, into the region $z > 0$.
As expected we find from
\begin{eqnarray}
\mbox{\boldmath$D$}^A (z)
= \mbox{\boldmath$P$}^A(z)
- {\rm LIM}
\left\{
- \mbox{\boldmath$\dot V$} \OO^\re
\left( \mbox{\boldmath$V$}^{(0)A} + \frac{\beta_0}{2} \1 \right)
- \mbox{\boldmath$g$} \OO^\re \mbox{\boldmath$V$}^{(0)A}
+ \mbox{\boldmath$V$}^{(0)A} \OO^\re \mbox{\boldmath$g$}
+ \mbox{\boldmath$G$}^{A}
\right\}
\end{eqnarray}
a simple convolution-type representation for ER-BL kernels which can be
immediately deduced from the forward results for singlet $QQ$-channel
\begin{eqnarray}
\label{D-QQ-o}
{^{QQ}\! D}^A
= C_F^2 \left[ D_F \right]_+
- C_F \frac{\beta_0}{2} \left[ D_\beta \right]_+
- C_F \left( C_F - \frac{C_A}{2} \right)
\left[ \frac{4}{3} {^{QQ} v} + 2\, {^{QQ} v}^a \right]_+
- 6\, C_F T_F N_f {^{QQ} v}^a,
\end{eqnarray}
where $D_F$, $D_\beta$ are given by Eqs.\ (\ref{DF-QQ}) and
(\ref{Dbeta-QQ}), respectively. The rest of channels is expressed as
\begin{eqnarray}
{^{QG} D}^A
&=& 3\, C_F T_F N_f
\left\{
{^{QQ} v}^a \OO^\re {^{QG} v}^a
- \frac{1}{2} {^{QG} v}^a
\right\} \\
&-& 2\, C_A T_F N_f
\left\{
3\, {^{QQ} v}^a \OO^\re {^{QG} v}^a
+ \left[ 1 + 2 \zeta (2) \right] {^{QG} v}^a
\right\}, \nonumber\\
{^{GQ} D}^A
&=& C_F^2
\left\{ \frac{1}{2}
\left[ {^{GG} v}^A \right]_+ \OO^\re {^{GQ} v}^a
- \frac{3}{2} {^{GQ} v}^a
\right\}
- C_F \frac{\beta_0}{2}
\left\{
{^{GQ} v}^a \OO^\re \left[ {^{QQ} v} \right]_+
- \frac{1}{6} {^{GQ} v}^a
\right\} , \\
&-& C_F C_A
\left\{
\frac{3}{2} \left[ {^{GG} v}^A \right]_+ \OO^\re {^{GQ} v}^c
+ \left[
2 \left[ {^{GG} v}^A \right]_+ - \frac{1}{2} {^{GG} v}^a
\right] \OO^\re {^{GQ} v}^a
- \left[ \frac{7}{3} - 2 \zeta (2) \right] {^{GQ} v}^a
\right\} , \nonumber\\
{^{GG} D}^A
&=& C_A^2
\left\{
\left[
\left[ {^{GG} v}^A \right]_+ + \frac{1}{2} {^{GG} v}^a
\right] \OO^\re {^{GG} v}^a
+ \frac{2}{3} \left[ {^{GG} v}^A \right]_+
- \frac{1}{4} {^{GG} v}^a - 2 \delta(x - y)
\right\} \\
&-& C_A \frac{\beta_0}{2}
\left\{
- \frac{1}{2} {^{GG} v}^a \OO^\re {^{GG} v}^a
+ \frac{5}{3} \left[ {^{GG} v}^A \right]_+
 + {^{GG} v}^a + 2 \delta(x - y)
\right\} \nonumber\\
&-& C_F T_F N_f
\left\{
{^{GG} v}^a \OO^\re {^{GG} v}^a
- {^{GG} v}^a + \delta(x - y)
\right\} , \nonumber
\end{eqnarray}
where we have introduced a new kernel
\begin{eqnarray}
\label{kernel-c}
{^{GQ} v^c}(x,y)=
\theta(y-x)  \frac{x^2}{y}\left(2 \bar{x}y-\bar{y} \right) -
\left\{x\to \bar{x} \atop y\to \bar{y}  \right\}.
\end{eqnarray}
These results provide us with the explicit parity odd singlet
evolution kernels.

\section{Conclusions.}

In this paper, we have presented a simple method for construction of
the exclusive evolution kernels in NLO from the knowledge of the
conformal anomalies and the available two-loop splitting functions.
The main task was, of course, the reconstruction of the diagonal part of
the kernel in the basis of Gegenbauer polynomials.

In the course of study we have established convolution-type formulae for
the bulk of contributing two-loop graphs with an exception of cross-ladder
diagrams. The complications which arise in the restoration of the latter
from the known forward kernels has been overcome making use of $\cN = 1$
supersymmetric constraints \cite{BelMulSch98}. The former feature suggests
that by disentangling the topology of corresponding diagrams, it might
allow for an effective and facilitated way of explicit calculation. One
may expect that this property persists for a subset of diagrams at higher
orders and can be used, e.g.\ for diagrammatical derivation of NNLO
splitting functions.

The details of the present formalism together with the flavour singlet
parity even case, where new subtleties appear, will be discussed elsewhere.

\vspace{1cm}

A.B. was supported by the Alexander von Humboldt Foundation.

\end{document}